\documentclass{aastex} 
\let\captionbox\relax
\usepackage{spr-astr-addons} 
\usepackage{url}
\usepackage{amsfonts}
\usepackage{bm}
\usepackage{color}
\usepackage{comment}
\usepackage{epstopdf}
\usepackage{gensymb}
\usepackage{graphicx}
\usepackage[caption = false]{subfig}

\begin{document}
%
\title{Exoplanet System {\em Kepler}-2 with comparisons to {\em Kepler}-1 and 13}

\shorttitle{Exoplanet system {\em Kepler}-2}
\shortauthors{<Rhodes et al.>}

\author{Michael~D.\ Rhodes\altaffilmark{1}}
\affil{Brigham Young University, Provo, Utah, USA} 
\and
\author{\c{C}a\u{g}lar P\"{u}sk\"{u}ll\"{u}\altaffilmark{2}}
\affil{\c{C}anakkale Onsekiz Mart University,\ \c{C}anakkale, Turkey}
\and
\author{Edwin\ Budding\altaffilmark{2,3,4,5}}
\affil{University of Canterbury, Christchurch, New Zealand}
\and
\author{Timothy~S.\ Banks\altaffilmark{6}}
\affil{Nielsen, Chicago, USA}

\email{tim.banks@nielsen.com} 


\altaffiltext{1}{Brigham Young University, Provo, UT 84602, USA}
\altaffiltext{2}{Dept.\ of Physics, \c{C}anakkale Onsekiz Mart University,\ TR-17020, \c{C}anakkale, Turkey}
\altaffiltext{3}{Dept.\ Physics \& Astronomy, University of Canterbury, New Zealand.}
\altaffiltext{4}{SCPS, Victoria University of Wellington, P.O. Box 600, Wellington, New Zealand}
\altaffiltext{5}{Carter Observatory, 40 Salamanca Rd, Wellington, New Zealand}
\altaffiltext{6}{Data Science, Nielsen, 200 W Jackson Blvd, Chicago, IL 60606, USA.
Email: tim.banks@nielsen.com  Tel:+1-847-284-4444}

\vspace{2mm} 



\begin{abstract}

We have carried out an intensive study of photometric (Kepler Mission)
and spectroscopic data on the system {\em Kepler}-2  (HAT-P-7A) using
the dedicated software {\sc WinFitter 6.4}. The mean individual data-point
error of the normalized flux values for this system is 0.00015, 
leading  to the model's specification for the mean reference flux to an
accuracy of $\sim$0.5 ppm. This testifies to the remarkably high accuracy of
the binned data-set, derived from over 1.8 million individual
observations. Spectroscopic data are reported with the similarly
high-accuracy radial velocity amplitude measure of $\sim$2 m s$^{-1}$.
The analysis includes discussion of the fitting quality and model
adequacy.

Our derived absolute parameters  for {\em Kepler}-2 are as follows:
$M_p$ (Jupiter)  1.80 $\pm$ 0.13; $R_{\star}$ 1.46 $\pm 0.08 \times
10^6$ km; $R_p$  1.15 $\pm 0.07 \times 10^5$ km.  These values imply
somewhat larger and less condensed bodies than previously catalogued,
but within reasonable error estimates of such literature parameters.

We find also tidal, reflection and Doppler effect parameters, showing
that the optimal model specification differs slightly from a `cleaned'
model that reduces the standard deviation of the $\sim$3600 binned light
curve points to less than  0.9 ppm.  We consider these slight
differences, making comparisons with the hot-jupiter systems {\em
Kepler}-1 {\color{black} (TrES-2)} and 13.

We confirm that the star's rotation axis must be shifted towards the
line of sight, though how closely depends on what rotation velocity is
adopted for the star. From joint analysis of the spectroscopic and
photometric data  we find an equatorial rotation speed of 11 $\pm$ 3 km
s$^{-1}$.

A slightly brighter region of the photosphere that distorts the transit
shape can be interpreted as an indication of the gravity effect at the
rotation pole; however we note that the geometry for this does not match the
spectroscopic result. We discuss this difference, rejecting the
possibility that a real shift in the position of the rotation axis in
the few years between the spectroscopic and photometric data-collection
times.  Alternative explanations are considered, but we conclude that
renewed detailed observations are required to help settle these
questions. 

\end{abstract}





\section{Introduction}

Different techniques in exoplanet discovery and research have been
selectively applied to, or alternatively have led to the characterization of, different types of object. The {\em Kepler}
Mission has detected a relatively large proportion of `hot jupiters',
while high-resolution spectroscopy has located many objects of
comparable mass but often further from the host star, spectroscopy being
less selective in what it can sample. Spectroscopy has, however, proved
to have highly significant applications to certain {\em Kepler}
targets --- as studied below. Neuh\"{a}user et al.\ (2011) have reviewed
the relevant general background.

Borucki et al.\ (2003) set out the aims of the original {\em Kepler}
Mission within the context of exoplanet research, while a comprehensive
early summary was that of Borucki et al.\ (2011). The Kepler Science
Center manages the interface between the scientific mission and the {\em
Kepler} data-using community. Importantly, data are freely and easily
available from the NASA Exoplanet Archive (NEA-website).

Rhodes \& Budding (2014) developed and applied the program {\sc
WinFitter 6.4}\footnote{The program was called {\sc WinKepler} in 2014.} to
the analysis of {\em Kepler} exoplanet light curves and found their
results to compare favourably with those of other research groups.  Occasional 
small differences in such results, and their significance, were
discussed by Budding et al.\ (2016), including the case of {\em
Kepler}-1.  Apparent peculiarities of {\em Kepler}-1 have been
previously remarked on (Kipping \& Spiegel, 2011) regarding the
exoplanet's low albedo and possible variations of other parameters. Such
effects are further discussed below. The fitting program {\sc
WinKepler} (Rhodes \& Budding, 2014), has been reworked and upgraded to
the version {\sc WinFitter} 6.4 (cf.\ Budding et al., 2018), used in the
present study. {\sc WinFitter} 6.4 includes an option for fitting the
radial velocity curve, including the effects of eclipses  and proximity,
in a parallel way to the photometric data-fitting.  This is pursued in
Section~3.

Regarding the data, it should be noticed that  the NEA operates an
initial `pre-search data conditioning' (PDC) process aimed at removing
systematic, but non-object-related, trends. The methods applied for this
have evolved with growing experience of the system (Twicken et al.,
2010; Smith et al., 2012; Stumpe et al., 2012). The resulting PDCSAP
fluxes are given in the output columns of the data worksheets released
by the NEA.  Different opinions have been voiced about this, but the
PDCSAP fluxes can be checked by users for their self-consistency within
their own programmes. This approach is in regular use by NEA data
analysts and is followed in the present study.

This paper discusses  optimised fittings of a suitable function to
selected NEA-sourced data-sets in terms of parameters that characterize
relevant physical effects. We concentrate on the hot-jupiter containing,
relatively well-known, systems {\em Kepler}-1
 {\color{black}(TrES-2, discovered by O'Donovan~{\em et al.}, 2006)}, 
 -2  {\color{black} (HAT-P-7, discovered by P\'{a}l~{\em et al.}, 2008)} and -13 
 {\color{black} (Borucki~{\em at al.}, 2011)}, with a new
{\sc WinFitter} analysis of {\em Kepler-2} being the main focus of the
present work.  We include comparisons of our results with previous
analyses of {\em Kepler}-1 and -13.

It should be demonstrated that specified parameters satisfy formal
determinacy conditions. At the same time, the adequacy of the {\sc
WinFitter} model to explain the effects needs to be examined. These
issues have been clarified in previous papers on {\em Kepler} exoplanet
light curves (cf.\ Budding et al., 2018). For this system we used the
transit derived parameters from Welsh et al. (2010) and the stellar
parameters derived through asteroseismology by Christensen-Dalsgaard et
al. (2010).

A significant point concerns increasing confidence in the published
properties of exoplanets arising from the cross-comparisons of results
from different investigations.


\section{Prior information}

In Table~1 we summarize the prior information available from the NEA or
literature sources (e.g.\ Batalha et al., 2013)\footnote{Parameters
listed on the NEA website are in a process of intermittent revision and
upgrading.} needed to reach a full set of results from the Kepler light
curves. In preparing this table we have taken into account parameters
listed by various other authors; including the sources listed in our
previous studies of {\em Kepler}-1 and -13, (Budding et al., 2016 \&
2018; as well as numerous other papers on {\em Kepler}-2, including
P\'{a}l et al., 2008; Winn et al., 2009; Narita et al., 2009; Welsh et
al., 2010; Christensen-Dalsgaard et al., 2010; Faigler et al., 2013;
Lund et al., 2014; Angerhausen et al., 2015; Esteves et al., 2015;
Masuda, 2015; and others).

Our Table~1 numbers contain rounded averages from these sources. Note,
though, that the main derivables from photometric analysis are not so
dependent on absolute parameter specification. Fittings to the transit
minima alone lead directly to relatively well-defined estimates for
$r_1$, $k$ and $i$ (the radius of the star expressed as a fraction of
the semi-major axis of the orbit, the ratio of planet to stellar radii,
and the orbital inclination).

With regard to representative values for the 
 planets' mean surface temperatures $T_p$, we can write (cf.\ Catling~\& Kasting, 2017)
\begin{equation}
T_p \approx T_{\star} (1 - A_B)^{1/4} \sqrt{r_1/2}  \,\,\,  ,
\end{equation}
where $T_{\star}$ is the star's effective surface temperature and $A_B$ is the Bond albedo
(= 0 for a `black body').\footnote{The Earth's mean 
surface temperature, estimated in this way, is  $\sim$280 K,
while the regular GISS measure is close to 287 K.}

\begin{figure*}[t]
\centering
\subfloat[]{\includegraphics[width=3in]{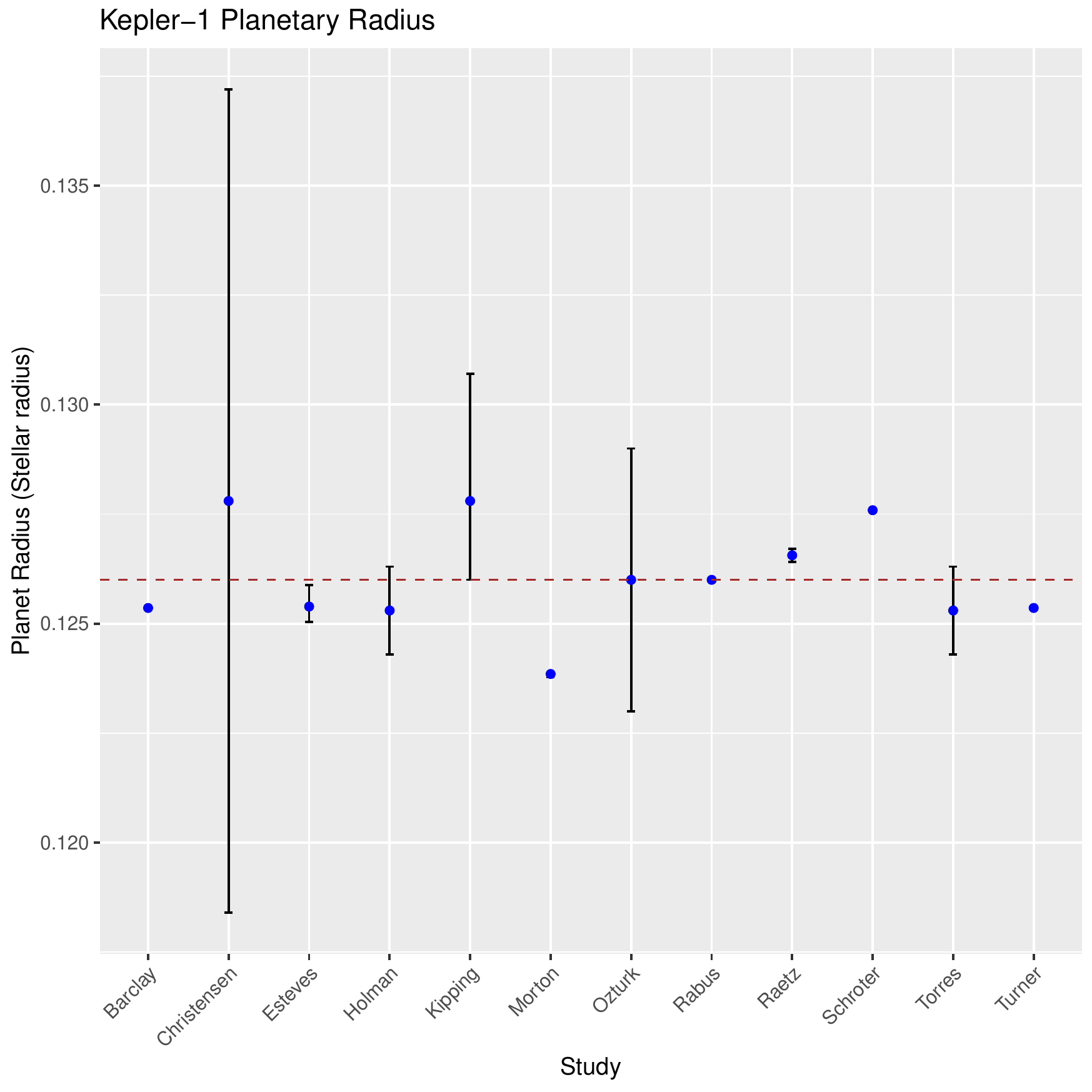}}
\subfloat[]{\includegraphics[width= 3in]{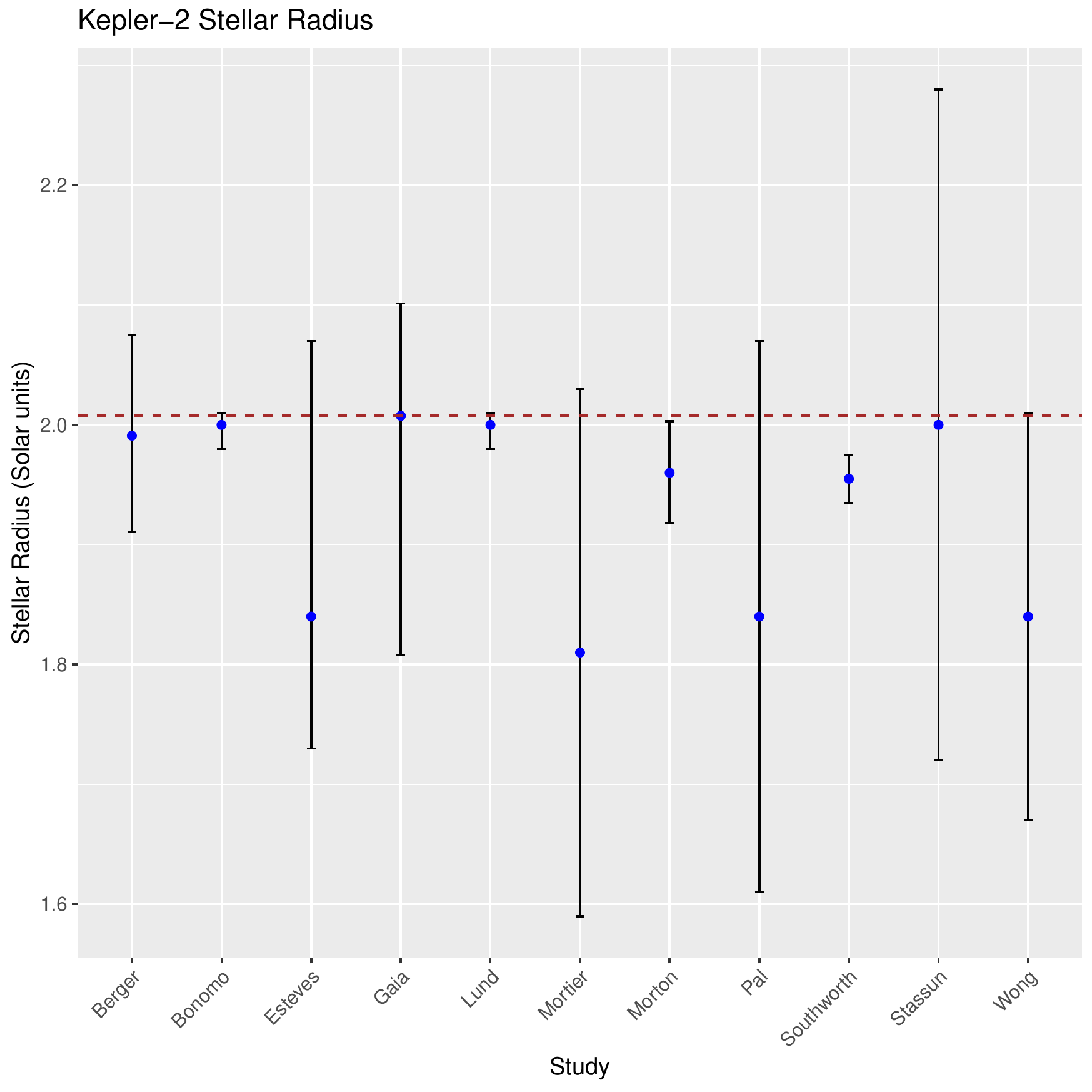}}\\
\caption{\color{black} {\bf Example Comparison of starting information:} 
sub-figure(a) shows the planetary radii taken from the NEA summary page for Kepler-1, and sub-figure (b) plots the stellar radii for Kepler-2.  The
initial starting estimates used by the current study are shown as brown dotted lines.  Papers are referred to by their lead author, with reference details
available on the appropriate NEA summary webpage.  The Gaia DR2 (Gaia Collaboration, 2018) is referred to as `Gaia'. Where Gaia data were available we used these in preference
to our mean values (as in sub-figure (b)).  These representative charts are to show that our starting estimates are in line with the literature, and also to demonstrate the wide spread
in error estimates across studies (which were taken as reported by the individual studies).}
\label{fig:one_kep}
\end{figure*}

\begin{table*}
\label{tbl-1a}
{\footnotesize
\begin{center}
\hspace{2em} \caption{Primary Input Data \label{tbl-1}}
\begin{tabular}{|r|r|r|r|r|r|r|r|r|}
\hline 
\multicolumn{1}{|c|}{KOI} & 
\multicolumn{1}{|c|}{$L_1$} &
\multicolumn{1}{|c|}{$M_*$} & 
\multicolumn{1}{|c|}{$R_*$} & 
\multicolumn{1}{|c|}{$T_*$} & 
\multicolumn{1}{|c|}{$T_p$} & 
\multicolumn{1}{|c|}{$\lambda$} &
\multicolumn{1}{|c|}{Epoch} & 
\multicolumn{1}{|c|}{$P$} \\ 
\hline 
1.01 & 0.9683 & 1.06 & 0.98 & 5850 & 1350& 6220 & 2454849.52664& 2.47061892 \\ 
2.01 & 1.000 & 1.56 & 2.01 & 6390 & 2000& 6140 & 2454954.358470& 2.204740  \\ 
13.01& 0.55   & 2.27 & 2.70 & 8500 & 2500 & 5970& 2455138.7439  & 1.7635877 \\ 
 \hline 
\end{tabular} 
\end{center}}
\vspace{1ex}
\begin{footnotesize}
\noindent
 The notation here is: $L_1$ -- fractional luminosity of host star taking into account 
 the companion (see Section 2.1), $M_*$ -- mass of star (solar masses), $R_*$ -- 
 radius of star (solar radii), $T_*$ -- temperature of star (K), $\lambda$ -- 
 effective wavelength ({\AA}), $P$ -- orbital period (in days).
The numbers given in this table are rounded averages from results in the cited 
literature and sources provided by the NEA (see text {\color{black} and also Figure~\ref{fig:one_kep}), bar where Gaia DR2 figures were available and therefore adopted}.
\end{footnotesize}
\end{table*} 

\setcounter{table}{0}
\begin{table*}
\label{tbl-1b}
{\footnotesize
\begin{center}
\hspace{2em} \caption{Primary Input Data (contd.)\label{tbl-1a}}
\begin{tabular}{|r|r|r|r|r|r|r|r|r|r|}
\hline 
\multicolumn{1}{|c|}{KOI} & 
\multicolumn{1}{|c|}{$\bar{\rho_{\star}}$ } & 
\multicolumn{1}{|c|}{$\bar{\rho_p}$ } & 
\multicolumn{1}{|c|}{$Z$} & 
\multicolumn{1}{|c|}{$\log g$} & 
\multicolumn{1}{|c|}{$a$} & 
\multicolumn{1}{|c|}{ ${M_p}/{M_*}$} & 
\multicolumn{1}{|c|}{ ${R_p}/{R_*}$} & 
\multicolumn{1}{|c|}{$u_1$} &
\multicolumn{1}{|c|}{$A_g$} \\ 
\hline 
1.01  & 1.44 & 0.90 &  --0.15  & 4.44 & 0.0353  & 0.0013  & 0.1260 & 0.64 & 0.03 \\ 
2.01  & 0.27 & 0.61 &  --0.15  & 4.07 & 0.03676 & 0.0012 & 0.081  & 0.48 & 0.27\\ 
13.01 & 0.16 & 1.00 &  --0.141 & 3.94 & 0.0349  & 0.0036  & 0.0855 & 0.44 & 0.61\\ 
 \hline 
\end{tabular} 
\end{center}}
\vspace{1ex}
\begin{footnotesize}
\noindent
 The notation is now: $\bar{\rho_{\star}}$  -- star's mean density (CGS), 
$\bar{\rho_p}$ -- planet's mean density,
 $Z$ -- metallicity of star, $\log g$ -- log$_{10}$ of the surface gravity of star (cgs units), $a$ -- semi-major axis in AUs, $M_p/M_*$ -- ratio of planet to star masses, $u_1$ -- stellar (linear) limb-darkening coefficient. The numbers given in this table are rounded averages from results in the cited literature and sources provided by the NEA (see text).
\end{footnotesize}
\end{table*} 


\subsection{{\em Kepler}-1}

{\em Kepler}-1 (= KOI 1.01; KIC 11446443)\footnote{KIC stands for Kepler
Input Catalogue, KOI for Kepler Object of Interest.} was first
designated TrES-2b by the Trans-Atlantic Exoplanet Survey (TrES) (Alonso
et al., 2004). This relatively bright system was intended to be within
the limited field of view of the original {\em Kepler} survey. Relevant
evaluations of its properties were published before the launch of the
{\em Kepler} spacecraft (O'Donovan et al., 2006; Sozzetti et al., 2007;
Winn et al., 2009). Derived mass and radius values indicated a gas giant
having a composition and structure similar to Jupiter, but relatively
close to its star. It thus became an early representative of the hot
jupiter type of exoplanet. The host star GSC 03549-02811 is at an
estimated distance of $\sim$250 pc (Sozzetti et al., 2007; Rhodes \&
Budding, 2014).

It was noted relatively early (Daemgen et al, 2009) that TrES-2b is in a
visual binary system: a point that needs to be taken account of in the
photometric analysis. O'Donovan et al.\ (2006) and Sozzetti et al.\
(2007) also took into account spectroscopic data. They derived fairly
consistent mass and radius values for the planet of about 1.2 M$_{\rm
Jup}$ and 1.2 R$_{\rm Jup}$, respectively. Further details can be found
in our previous study (Budding et al., 2016).

 
\subsection{{\em Kepler}-2}

This exoplanet system is also known as HAT-P-7Ab, 2MASS
J19285935+4758102, TYC 3547-01402-1b, KIC 10666592 and BD+47 2846b.  The
`b' suffix on the BD number shows that  {\em Kepler}-2 is also in a
visual binary system. However, the separation in Kepler's camera is
about 30 pixels; sufficiently distant to prevent significant image
contamination from the companion (Dotson, 2013; Libralato et al., 2016).
The planet's 10.5 mag host star has been associated with a late A to
mid-F type classification: a Main Sequence object situated  at about 350
pc distance. Its `hot jupiter' companion, that produces a $\sim$0.67\%
variation of light, was also discovered before the launch of the Kepler
satellite in 2009 (P\'{a}l et al., 2008) and the system has figured in
numerous studies and survey articles over the last decade.

Measurements attributed to the Schlesinger-Rossiter-McLaughlin
effect\footnote{Often just referred to as the  Rossiter effect.}
{\color{black}(Schlesinger, 1910; Rossiter, 1924; McLaughlin, 1924)}
suggested a high angle ($\gtrsim 90\degr$) between the stellar equator
and the planetary orbit (Winn et al., 2009; Narita et al., 2009;
Albrecht et al., 2012). The implied retrograde motion  (the first such
example, cf.\ Narita et al., 2009) is awkward to explain directly from
the classical `nebular hypothesis' of planet system origins: something
additional is sought. Planet-planet scattering, or a Kozai effect, may
be involved though only one planet has been definitely identified in the
system so far.

Welsh et al.\ (2010) announced their discovery of tidal effects in the
light curve, the first exoplanet light curve to be so characterized.
Borucki et al.\ (2009) reported the light curve to be consistent with a
strongly absorbing atmosphere above the planet, but with limited heat
transport to the night side.  On the other hand, Esteves et al.\ (2015)
drew attention to changes of phase  of the system's peak brightness,
possibly arising from inhomogeneous clouds and/or substantial
atmospheric winds. They related these to  albedo values and the planet's
overall energy budget, although the system was not included with the
`super-rotation' candidates listed by Angerhausen et al.\ (2015).

Alternative ideas on brightness anomalies were initiated by Morris et
al.\ (2013), who considered a  spot-like gravity darkening effect.  Van
Eylen et al.\ (2013), studying a larger data sample, cautioned about the
possibility of instrumental artefacts, a noteworthy point in the context
of {\em Kepler}-2's known visual companion (see also, Lund et al.,
2014). Masuda (2015)  applied gravity-effect analysis to the complete
data-set, finding general support for a near-pole-on configuration.

Masuda (2015) also referred to  the previous work of Benomar et al.\
(2014) and Lund et al.\ (2014) on {\em Kepler}-2, that included detailed
discussions of the results of asteroseismology. Due to the  very low
noise levels and uninterrupted long-term data monitoring with short
sampling cadences, satellite-based photometry may allow
asteroseismological techniques to recover basic stellar parameters with
relatively high precision. Further background on such applications can
be found in the collection of papers from Shibahashi \& Lynas-Gray
(2013), Guzik et al.\ (2014) and others. Even greater information yield,
or confidence on parameter estimates, can be gained when such
information is combined with other  forms of data, or approaches to
analysis (e.g., Lillo-Box et al, 2016).
 
Benomar et al.\ (2014)  noted that the apparent profiles of rotationally
induced frequency multiplets are sensitive to the angle between the line
of sight and the star's rotation axis (usually denoted $i_{\star}$),
permitting a good estimate of its value from asteroseismology. However,
differences exist between parameters from different investigations using
these techniques. For example, according to Benomar et al.\ (2014) the
host-star's mass ($M_{\star}$ in solar units for this and the following)
is 1.59$\pm$0.03; Lund et al.\ (2014) have it at 1.51$\pm$0.04 or
1.63$\pm$0.09, using different models; van Eylen et al.\ (2013) find
1.36$\pm$0.02;  while Christensen-Dalsgaard et al.\ (2012) report
1.52$\pm$0.04. The latter is close to the average value of these
estimates, though the standard deviation is about 7\% of that mass.

More recently, Wong et al.\ (2016) examined infra-red data on {\em
Kepler}-2 from the Spitzer Space Telescope.  Their measured depths of
the secondary eclipses were consistent with a relatively high daytime
representative temperature of $\sim$2650. With the aid of detailed
atmospheric modelling, their analysis  pointed to a relatively
inefficient day-night heat circulation for {\em Kepler}-2. An eastwardly
shifted `hotspot', which has been associated with a super-rotating
equatorial jet in comparable studies of this and other hot jupiters
(Faigler et al., 2013; Faigler \& Mazeh, 2015), was not detected by Wong
et al.\ (2016), though the models they used were not consistent with all
aspects of the data.

On the other hand, Armstrong et al (2016) repeated earlier findings of
shifts of peak brightness, and suggested recurrent movements of a hot
spot from one side the of the planet's substellar point to the other on
timescales of tens to hundreds of days. Their interpretation involved
significant variations in global wind patterns and related cloud
coverage.

The distribution of individual observations, for example in Fig~2
of Armstrong et al.\ (2016), shows that the proportion of points whose
error bars are completely missed by the fitted model curves is much
greater than might be expected. In other words, systematic short-term effects are
present in the data that are unaccounted for by generic models for the
complete phase-range. Precipitation of crystallised alumina has been
proposed to explain the clouding in the hot conditions experienced by
this exoplanet, leading to a picturesque description of ``raining rubies
and sapphires'' in popular reports {\color{black} (see Armstrong~{\em et al.}, 2016,
for the source paper and Dvorsky (2016) for an example of a popular report).}

Given the possibility of over-detailed modelling, leading to parameter
indeterminacy, a relatively simple summary of clouding effects was
proposed by von Paris et al.\ (2016) on the basis that they would cause 
asymmetric primary transits due to a predominance of cloud towards the
western (`morning') limb of the planet (the leading limb in primary
transit). The eastern limb, which becomes visible as the planet recedes
after mid-transit, is relatively cloud-free in this scenario.  There has
been a common assumption of super-rotating winds above a rotationally
synchronized planet body in cloud-formation discussions. Clouds formed
to a greater extent on the unilluminated hemisphere are then transported
to the morning sector that would become more visible on approach. In
view of an expected high opacity at the cloud tops, the planet could
have different apparent radii for the two sides of the eclipse.
  
Still, ambiguity remains about the possible causes of minimum asymmetry
and light curve anomalies.


\subsection{{\em Kepler}-13} 

The near-10th magnitude exoplanet-containing system {\em Kepler}-13 was
identified early in the mission's output (Borucki et al., 2011) and
given the designation KIC 9941662 (Brown et al., 2011) after the first
two quarters of data collection (see also Rowe et al., 2011). This
complex, massive hot jupiter has an orbital period of $\sim$1.764 d and
an (asymmetric) transit depth of about 0.45\% of full light. Numerous
papers have been presented on the system, as briefly summarized in our
previous contribution (Budding et al., 2018).

Rhodes \& Budding (2014) remarked on KOI-13.01 as of special interest
from even early studies, since  proximity effects and a secondary minimum are
noticeable even from preliminary inspection. These are not commonly
observed features of exoplanet light-curves (Esteves et al., 2013).
Rhodes \& Budding essentially confirmed the mass ratio of Mislis \&
Hodgkin (2012), while their value for the planet's mass (9.4 $\pm$1.4
M$_J$) was somewhat closer to the estimate of Mazeh et al.\ (2012).

Barnes et al.\ (2011) first called attention to the fact that photometry
of {\em Kepler}-13 may be compromised by the relatively bright companion
star, entailing slight shifts of the light centre on the detection array
affecting the data. Basically,  the aperture was set too small for the
first two quarters of data-acquisition, and so data from those quarters
have been subsequently disregarded. Having then restricted their
attention to the later short-cadence photometry, Barnes et al.\ (2011)
interpreted apparent variations in the detailed shape of the light curve
minima as arising from a gravity-darkened, rapidly rotating, host star,
scanned by the planet's transit with particular spin-orbit
configurations relative to the line of sight.

Numerical details in the model of Barnes et al.\ (2011) were challenged
by Johnson et al.\ (2014) who used a spectrographic, Doppler-tomographic
technique that indicated KOI-13.01 has a different orbit orientation to
that given by Barnes et al.\ (2011).  But interdependence of modelling
parameters in this kind of data-analysis should be kept in mind. The
planetary path followed in the model of Barnes et al.\ (2011) may need
to pass closer to the pole, increasing the relevant tilt-angle
($\lambda$), if that pole were not as bright. Hence, the relative scale
of an adjustable gravity-darkening could be weaker and the model of
Barnes et al.\ still hold, but with a different angle of tilt.

Mislis  \& Hodgkin (2012) confirmed that the planet orbits the slightly
brighter of the two stars (cf.\ Santerne et al., 2012) and also
confirmed that it has its own small, but detectable, light flux in the
broad, near-IR-containing, Kepler passband.

 
\section{New study of Kepler-2}

We continue along the lines of our  previous papers on {\em Kepler}-1
and -13 in dealing with the data on {\em Kepler}-2. 

\subsection{Preliminary photometric analysis}

\begin{figure}[t]
\includegraphics[width=8cm]{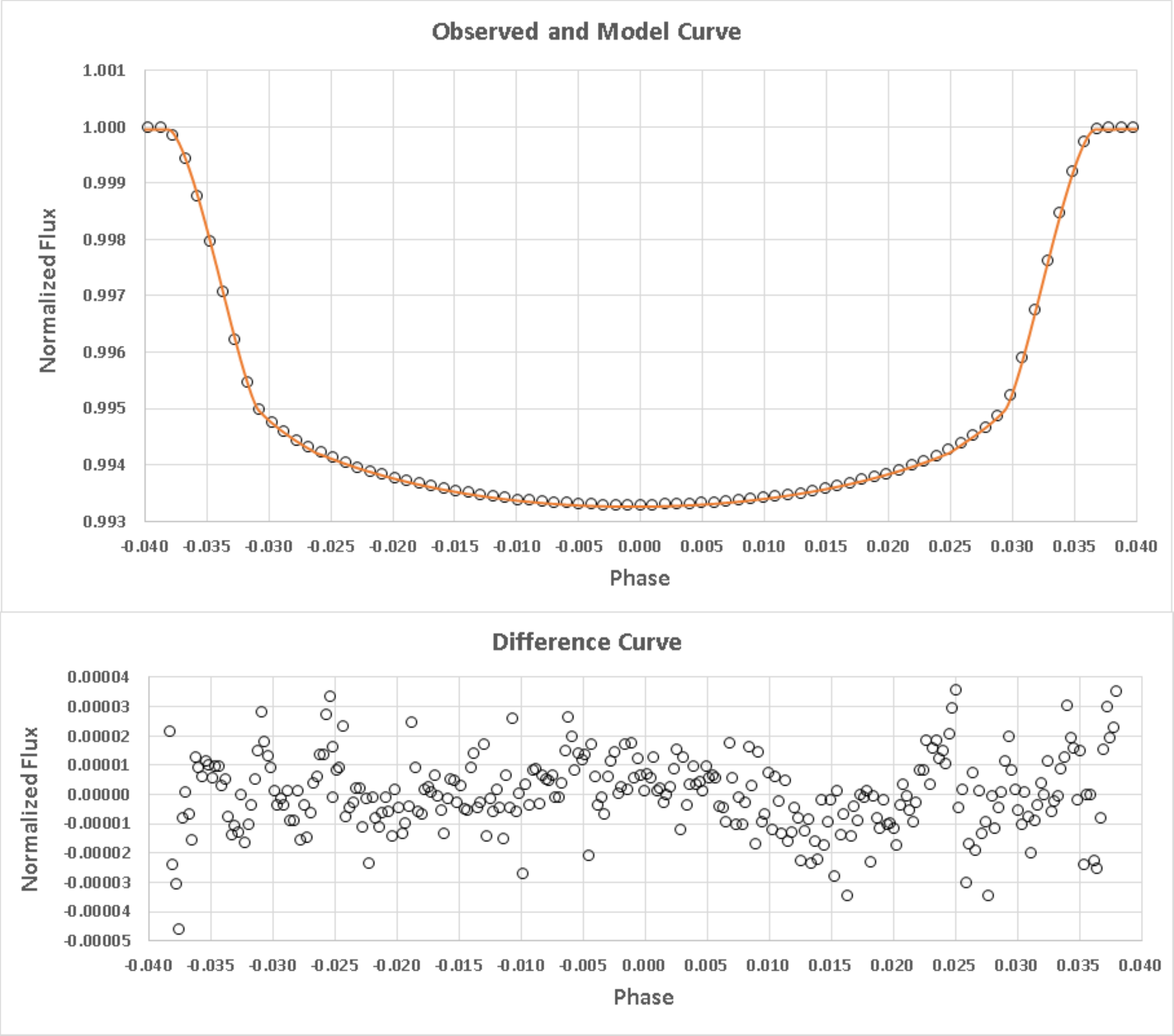} 
\label{fig:Kep2transit2}
\caption{The transit region for  
 {\em Kepler}-2b, matched by a basic {\sc WinFitter} model.  
The fitting-function will allow shifts of the rotation axis position and
rotation rate, and a quadratic  limb-darkening prescription,
but these are not used in the initial fitting.
 The corresponding residuals are shown in the lower panel. 
A systematic drop of about 0.00002 of the mean flux level, centred
on phase $\sim0.015$ can be noticed. This could be interpreted as 
the planet passing over a brightened region of surface:
feasibly a gravity-brightened polar region of the star.
} 
\end{figure} 
 
We used the program {\sc WinFitter 6.4} (Rhodes, 2019) to analyse the
available normalized PDCSAP short cadence data sets covering the
observing quarters 0-17, downloaded from the NASA Exoplanet Archive. The
given times of observation (BKJD) were converted to phases in the range
0.0 to 1.0, using the ephemeris given in Table 1. This produced a total
of 1,879,376 individual observations for the initial data-set. {\sc
WinFitter} has a binning application, which was used to bin the data-set
with a 500 to 1 ratio down to 3758 individual points. This enabled a
faster exploration of data-space, with negligible loss of mean parameter
validity. {\color{black}  An error measure is published with each flux measurement  on the NEA data-source.
We assume Poissonian statistics and infer a binned error measure by dividing the original by $\sqrt{N}$
where $N$ is the number of points in the bin.}  From this complete light curve, we extracted 374 data points
covering the phase region --0.05 to +0.05 to enable detailed analysis of
the transit.

The downloaded PDCSAP fluxes have been preliminarily normalized to
unity, but the process of optimizing the model-fitting allows for slight
variations of this quantity. For exoplanet light curves, we set, at
least initially,  the reference light level $U = L_1$ (fractional light
of eclipsed star). The transit region has a relatively large information
content per datum for the main geometric and limb-darkening parameters
($r_1$, $k$, $i$, and $u_1$). Having found good estimates for these
parameters, the more slowly varying, low-amplitude out-of-eclipse flux
changes can be used more effectively to bear on other parameters ---
such as the mass ratio, gravity-effect, or reflection coefficients ($q$,
$\tau$, $E$). 

 {\color{black} Background to the  program {\sc WinFitter}\footnote{The latest version of {\sc WinFitter}
is downloadable from {\bf {\footnotesize http://michaelrhodesbyu.weebly.com/astronomy.html }}} was presented in some 
 detail by Budding et al.\ (2018).
The current version is {\sc WinFitter 6.4}. It continues to be checked and upgraded when necessary.
{\sc WinFitter} uses a modified Marquardt-Levenberg technique to perform its
optimization procedure.
 Its fitting function is based on the {\em Radau} model developed from Kopal's (1959) presentation of the
 tidal and rotational distortions (`ellipticity'), and an appropriate description of the radiative interactions (`reflection').  
The photometric Doppler effect (Hills \& Dale 1974; Shporer et al., 2012) is also
included in the calculations. 
The algebraic form of the fitting function allows large regions of the $\chi^2$-parameter hyperspace to be explored at low computing cost.
  
Optimal models produced by {\sc WinFitter} correspond to the least value of $\chi^2$, defined as $\Sigma(l_{o,i} - l_{c,i})^2/\Delta l_i ^2$ (Bevington, 1969), 
where $l_{o,i}$ and $l_{c,i}$ are observed and calculated light levels at a particular phase.  $\Delta l_i$ is an error estimate for the measured values of $l_{o,i}$. 
The NASA Exoplanet Archive (NEA) lists empirical values of $\Delta l_i$ for each datum, whose mean allows a suitable setting for the mean datum error $\Delta l$
in {\sc WinFitter}.   The $\chi^2$ Hessian is numerically evaluated in the vicinity of the $\chi^2$ minimum.  Inspecting this matrix, in particular its eigenvalues and eigenvectors, 
gives insight into parameter determinacy and interdependence.  The Hessian is inverted to yield the error matrix.  
This must be positive definite if a determinate optimal solution is to be evaluated.  Application of this provision entails that over-fitting of the data is avoided.
 
For further information on {\sc WinFitter} and related issues see Budding et al.\ (2018).
Relevant early papers are Budding (1974) and Banks~\& Budding (1990). 
 }
 
The results of the initial transit analysis are summarized in Fig~2 and
Table~2. Adopted bolometric gravity-darkening and luminous efficiency
coefficients are listed in this table. These are converted to the values
applying at particular wavelength and effective temperature combinations
using the  black body approximations specified by formulae 9.18 and 9.19
in Budding \& Demircan (2007) using relevant data from Table~1. These
coefficients have relatively little effect for the primary transit
region of the light curve, though they will be  required in subsequent
analysis. In practice, at least in the first approximation to exoplanet
light curves, it is only the star's gravity effect $\tau_1$ and the
planet's reflection coefficient $E_2$ that are significant.

The shift from the nominal zero point to recover optimal phasing $\Delta
\phi_0$ in Table~2 has the rather large value of 0.233$^{\circ}$ or
0.00065 in phase. This can be essentially attributed to our use of an
earlier NEA epoch of BKJD 121.3578, rather than the revised value in
Table~1. That is equivalent to BKJD 121.35847, which will give slightly
later phases than those originally adopted. The period estimate has been
maintained throughout the 4 year data interval.

In Table~2 it is seen that two linear limb-darkening coefficients have
been used. A law which is linear in the cosine of the angle of
foreshortening appears adequate for the range of annular phases of the
transit ($u_1$[a]).  At the extreme limb, corresponding to partial
eclipses, however, it is well-known that the darkening becomes complex
and the simple linear law inadequate for precise modelling 
of the radiation transfer (Kourganoff, 1952). 
There will, however, be some  representative linear coefficient that can
optimally reduce the residuals in the short phase range towards the limb
($u_1$[p]), and though the residuals' scatter can be seen to increase
around the partial phases in Fig~2 it is not changed drastically. 
{\color{black} The
device of using a second mean linear coefficient to cover the short range of partial 
phases  is regarded as an artifice, brought about by the complexity of
the limb-darkening effect at the extreme limb.  This rather coarse
approximation is reflected in the somewhat greater variation of residuals 
through the partial phases, but the net effect on the other fitting parameters
is small, while details of the flux emanating from the
extreme limb region are not relevant to the main targets of our study. }
 
\begin{table}[h]
\begin{center}
\caption{Initial curve fitting results for the complete set of {\em Kepler} 
photometry, binned by a factor of  500, for the transit of {\em Kepler}-2b. 
Parameters have their conventional 
designations 
 as $U$: the `unit of light';  
$\Delta \phi_0$: shift of zero point of the phase scale;
$r_1$: radius of host star in units of the separation
of the two mass centres; $k$: ratio of the radii; $i$: orbital inclination; 
$u_1$ coefficient of the linear term in the limb-darkening law -- see Section (3.1).  
 $\chi^2/\nu$ : reduced $\chi^2$-value; $\Delta l$ : adopted mean error of a single datum. 
 The error estimates, applying to adjustable parameters,
  are derived from the formal error matrix calculated at the adopted optimum.
 Other fitting-function parameters are discussed in later sections of the paper.
\label{tbl-2}} 
\begin{tabular}{lcc}
\hline
\multicolumn{1}{c}{Parameter} & \multicolumn{1}{c}{Value} & 
\multicolumn{1}{c}{Error} \\
\hline\\
$U$ & 0.999992& 0.000002 \\    
$\Delta\phi_0$ (deg) & 0.233 & 0.021 \\   
$r_1$ & 0.242 & 0.0004 \\ 
$k$ & 0.0781 & 0.0001 \\  
$i$ (deg) & 83.08 & 0.01 \\  
$u_1$[a] & 0.457  &0.002 \\  
$u_1$[p] & 0.481 &0.008 \\
 $\tau$        & 1         &   ---     \\  
 $E$           & 1            &   ---    \\    
$\chi^2/\nu$ & 0.98 & \\   
$\Delta l $& 0.000011 & \\  
\hline
\end{tabular}
\end{center}
\end{table}
 
The adopted single point error is down on the raw datum error (0.00015; 
Budding and Rhodes, 2014) by a factor of 13.6, but this should be
$\sqrt{500} = 22.4$, if the scatter was due only to Poissonian counting
noise. It was mentioned already (Section 2.2) that additional sources of
deviation, perhaps related to chaotically variable cloud cover, affect
the data for {\em Kepler}-2. However, Table~2 reflects only a
preliminary fitting, and there may remain other systematic effects in
the data that {\sc WinFitter}'s full modelling parameter set can account
for.

Thus, the residuals plotted in Fig~2 show a small, but systematic,
depression centred around phase 0.015, that could be associated with the
eclipse of a brighter than average region of surface.  If this were a
polar region, brightened by the gravity-darkening effect, it should
inform about the rotational geometry, as in the case of {\em Kepler}-13.
Information on the rotation also derives from the the spectroscopy of
{\em Kepler}-2 that we examine next.


\subsection{Radial velocity curve}
 
{\em Kepler}-2 has been observed with the High Resolution Spectrograph
(HIRES) on the Keck I 10 m telescope, as well as the High Dispersion
Spectrograph (HDS) on the Subaru 8 m telescope, both instruments at the
high altitude Mauna Kea location. Initially, 8 HIRES spectra were
gathered by P\'{a}l et al.\ (2008). A further 9 were added by Winn et
al.\ (2009). All but one of these HIRES spectra were from the
out-of-eclipse phase range. Some 49 HDS spectra were obtained in June
and July, 2009 including a thorough covering of the planetary transit.

Observational details were reported by the foregoing authors, as well as
Narita et al.\ (2009). Essentially, the methods followed to obtain
radial velocities (RVs) were the same as those of the  California Planet
Search (Howard et al., 2009). It may be noted, however, that P\'{a}l et
al.\ (2008) refer to Butler et al.\ (1996) as the basis of their
reductional procedure, while Winn et al.\ (2009) cite Butler et al.\ (2006)
``with subsequent improvements''. The latter reference mentions having
introduced considerable revisions to methodology after their earlier
work, including a more accurate barycentric correction, as well as
changes to hardware that restricted previous precision. Moreover,
exoplanet RVs were originally arranged so that the zero point was set to
a median value of the RV measures.  In the later reference, however, an
offset has been fixed after an orbital solution for the system's mean
radial velocity $V_\gamma$. Butler et al.\ (2006) indicate that
differences may arise in the reference velocity for the exoplanet RV
curve due to these procedural changes, as well as from the binning of
groups of individual measurements. These  points are relevant, because
Winn et al.\ (2009) comment on a systematic difference in the reference
velocity ($V_{\gamma}$ in Table~3) between their RV curve and that of
P\'{a}l et al.\ (2008), that we look into below. 
 
We adopted the 86 RV values listed by Winn et al.\ (2009) as the basis
of an optimal model fitting using the equivalent (`Radau') model used
for our photometric analysis but tailored to RV measures. The model
accounts for the Doppler effects arising from a Keplerian orbit, but
having the spectral line positions corresponding to a centre of light
modified by the regular proximity effects that operate in close binary
systems (cf.\ Kopal, 1959, Ch.\ 5). An eccentric orbit was trialled in
the RV-curve fitting, but eccentricity parameters  could not be resolved
from zero (i.e.\ a circular orbit) with sufficient confidence from the
limited phase coverage.

Table~3 presents the results of fitting a rotating star model to the
spectral line shifts, with the assumption that the stellar rotation is
typical for Main Sequence stars of this mid-F type.  For such a star an
equatorial rotation velocity of 30 km s$^{-1}$ could  be reasonable
(McNally, 1965). Synchronised rotation of the star to the orbital motion
would give the relatively high speed of $\sim$46 km s$^{-1}$ at the
star's equator, hence its rotation parameter $\gamma_1$\footnote{The
ratio of the $i$-th component's angular velocity of rotation to that of
orbital revolution is denoted $\gamma_i$ in Kopal's (1959) book (Eqn
5-11 in Ch.\ 2), though in Budding et al.\ (2018) it was identified as
$\sqrt{\kappa}$.} is here set to 0.65.

The corresponding data and model curve are displayed in Fig~3.  An
inherent ambiguity arises, in that the RV measures only refer to motion
in the line of sight.  We have located the vector corresponding to a
positive rotation when viewed from above, i.e.\ the `north pole', this
being in the third quadrant of the stellar disk as observed at transit.
But the same line of sight effect can be produced by putting the
negative rotation pole (`south') mirrored in the first quadrant, with
the planet moving in the opposite direction.

\begin{figure}[t]
\includegraphics[width=8cm]{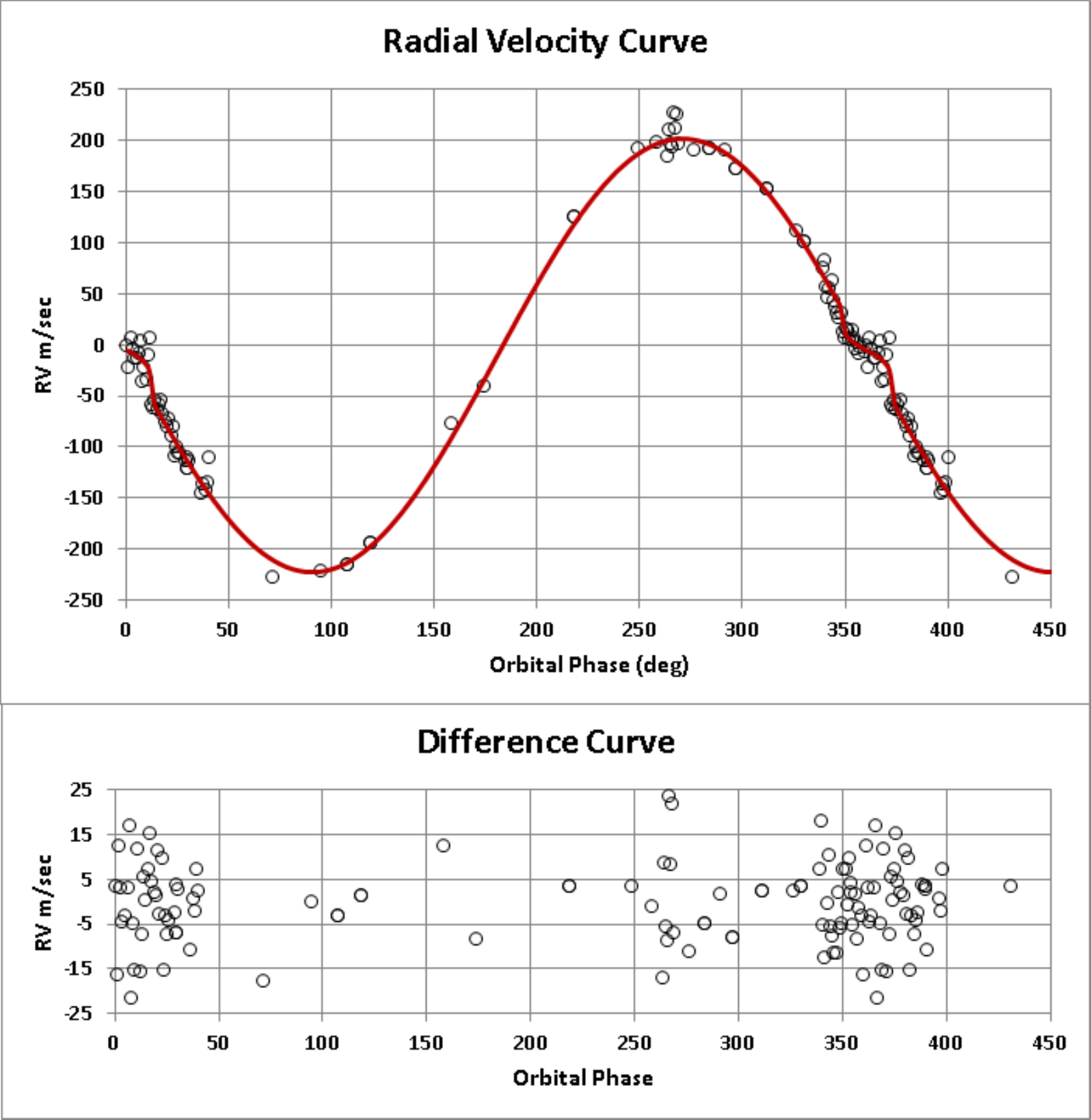}
\label{fig:Kep2rv}
\caption{Optimal fitting to the RV curve of Kepler~2A, using the full data-set cited by Winn et al.\ (2009).
The curve includes an optimized Rossiter effect modelling (Section 3.3),
but the sign an amplitude of this is atypical for binary system orbits.}
\end{figure} 
 
\begin{table}[h]
\begin{center}
\caption{Curve fitting results for the RV measures of Kepler~2A. 
Parameters have their conventional designations  as  in Table~1 but now with the addition of 
$K_1$: amplitude of primary's radial velocity variation; 
$V_{\gamma}$: mean radial velocity of the star
relative to the adopted reference frame  (see text); 
$\gamma_1$: ratio of star's angular velocities of rotation to orbital revolution;
$\lambda$: projected angle of tilt of the rotation axis;
$\Delta v$: adopted mean error of a single datum. The error estimates are 
derived from the formal error matrix calculated at the adopted optimum.  
Other parameters are passed from Table~2. 
The determined velocities are in m s$^{-1}$.
\label{tbl-3}
} 
\begin{tabular}{lrc}
\hline
\multicolumn{1}{c}{Parameter} & \multicolumn{1}{c}{Value} & 
\multicolumn{1}{c}{Error} \\
\hline\\
$K_1$  &      212.6    &   2.1        \\  
$\Delta \phi_0$ &  --0.5 &   0.3        \\
$V_{\gamma}$    & --8.6  &   1.4        \\
$i$   &       83.1      &              \\
$r_1$  &        0.242   &       \\  
$ r_2$ &       0.020    &       \\
$q$    &      0.0011    &       \\
$u1$   &       0.46     &       \\
$L_1 $  &       1.0     &       \\
$L_2 $   &      0.0     &       \\
$\gamma_1$  &     0.65
     &        \\  
$\epsilon$ &    93.8    &  1.3         \\
$\psi$    &     --1.2    & 1.2         \\
$\lambda$ &    188.8       & 1.2            \\
$\chi^2/\nu$ & 1.05     &       \\
$\Delta v$ &    10.4         &       \\
\hline
\end{tabular}
\end{center}
\end{table} 
  
Note that the polar axis points quite close to the line of sight in this
fitting (cf. Winn et al., 2009), but this is dependent on the assumed
scale of the angular velocities of rotation and revolution, as a
consequence of the Rossiter effect.  This was mentioned above and is
further discussed in the next subsection. If the star's rotation rate
parameter $\gamma_1$ is reduced from its adopted value, the pole can
move further from the line of sight into the third quadrant of the
Earth-facing disk (see Section~3.3).

For the most part, the out-of-eclipse RVs follow the near-sinusoidal
trend expected of the star's slight response to the orbital pull of the
planet. However, Winn et al. (2009) noted that there was a systematic
shift between the two sets of data gathered in mid-2007 and mid-2009 of
some 22 m s$^{-1}$ y$^{-1}$. We checked and used this correction in
modelling the combined data-sets.

During the transit phase range we encounter the Rossiter effect, which
permits a more detailed characterization of the system's properties. But
in the present case, the effect appears slight and lower than
could be generally expected. In fact, the departures from one continuous
sine curve for the complete data-set, given the accepted observational
errors of a few m s$^{-1}$, are sufficiently small as to allow
questioning the significance of {\em any} eclipse effect on the RV
measurements  (see Fig~3).
 
 We will return to this point: for now, we remark that
 given the prior mass of the star $M_{\star}$ from Table~1, the mass ratio $q$ can be 
 determined from a recursive solution of the implicit equation 
\begin{equation} 
 q = f^{1/3}(1+q)^{2/3}  \,\,\,  ,
\end{equation}
where the mass-function $f$ is given by
\begin{equation} 
f = C(1-e^2)^{3/2}K_1^3P/(M_{\star} \sin^3 i) \,\,\, .
\end{equation}

The values of period $P$ and inclination $i$ needed to solve Eqn(3) are
given in Table~1. The velocity amplitude $K_1$ can be read from the
results of the optimal parametrization the RV curve given in Table~3,
with the eccentricity $e$ being negligible. A recent value of the constant $C$
is  $1.03615 \times 10^{-7}$ (with $M_{\star}$ in solar units, $P$ in d
and $K_1$ in km  s$^{-1}$) to a sufficient accuracy (Budding, 2018),
which leads to $q = 0.00113$.  

As mentioned above, Winn et al (2009) suggested that this could produce
a light travel time effect in a bound system with a long enough period.
The shift in question  implies a drift from regular repetition in the
times of minima of order $3\times 10^8/c$ s (where $c$ is the velocity
of light),  i.e., seconds only. The period of the system in Table~1 is
given to this order of accuracy, or slightly better, (i.e.\ 6
significant digits after the decimal point for the period in d). It
therefore seems feasible to test if any such drift could be detected in
the run of $\Delta \phi_0$ values in a quarter-by quarter analysis of
the 17 relevant binned light curves.

Results of checking around this point are shown in Figs~4 and 5. While
long-term trends in the fiducial parameters $\Delta \phi_0$ and $U$ seem
vaguely possible, they can hardly be regarded as significant. Thus, a
net advance of $\sim$2 s, i.e. a slight tendency for the phase of
mid-minimum to {\em decrease} below zero, in the time of mid-transit
over the 17 quarters could be indicated in Fig~5.  Winn et al.'s (2009)
finding was, however, that the later radial velocities appeared to be
affected by an {\em increase} in the $\gamma$-velocity of the system,
i.e.\ a recession behind a regular sequence, though the corresponding
data-sets here are separated by a few years.

At the same time, the suggestion of a trend in Fig~5, not greater than
the scatter of individual results, but implying a variation in the mean
flux from the star of $\sim 10^{-5}$ over the $\sim$4 y period of the
Kepler mission is not backed by any decisive alternative evidence. So, while
there could be some slight indication of $\gamma$-velocity variation in
the observed times of mid-transit, or other, intermediate-term
photometric variations, no systematic effects of this kind can be
confidently confirmed from the available data.
 
Nevertheless, the rather large scale of random variations in these
fiducial parameters may have a connection with the discussion of
Armstrong et al.\ (2016),  referred to in Section~2.2.  We will return
to this in Section~4.
  
 \begin{figure}[t]
\includegraphics[width=8cm]{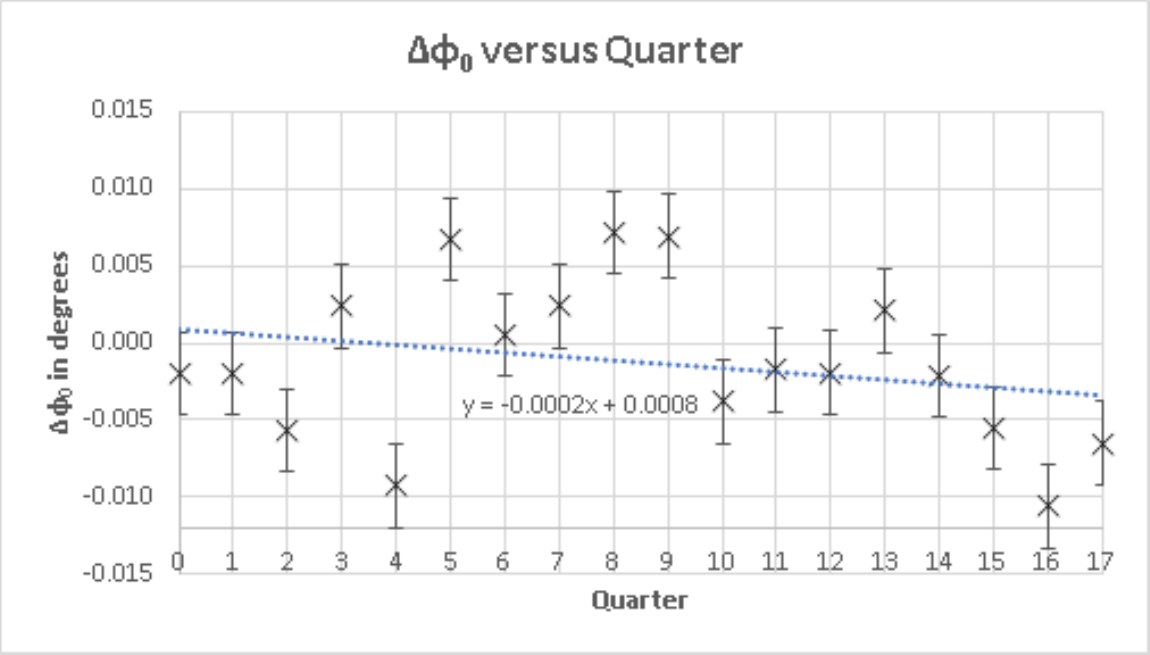} 
\label{fig:K2du}
\caption{Trend of correction to predicted mid-transit phases  (the parameter $\Delta\phi_0$) over
the 17 quarters of the complete data-set for Kepler 2.}

\end{figure}  
\begin{figure}[t]
\includegraphics[width=8cm]{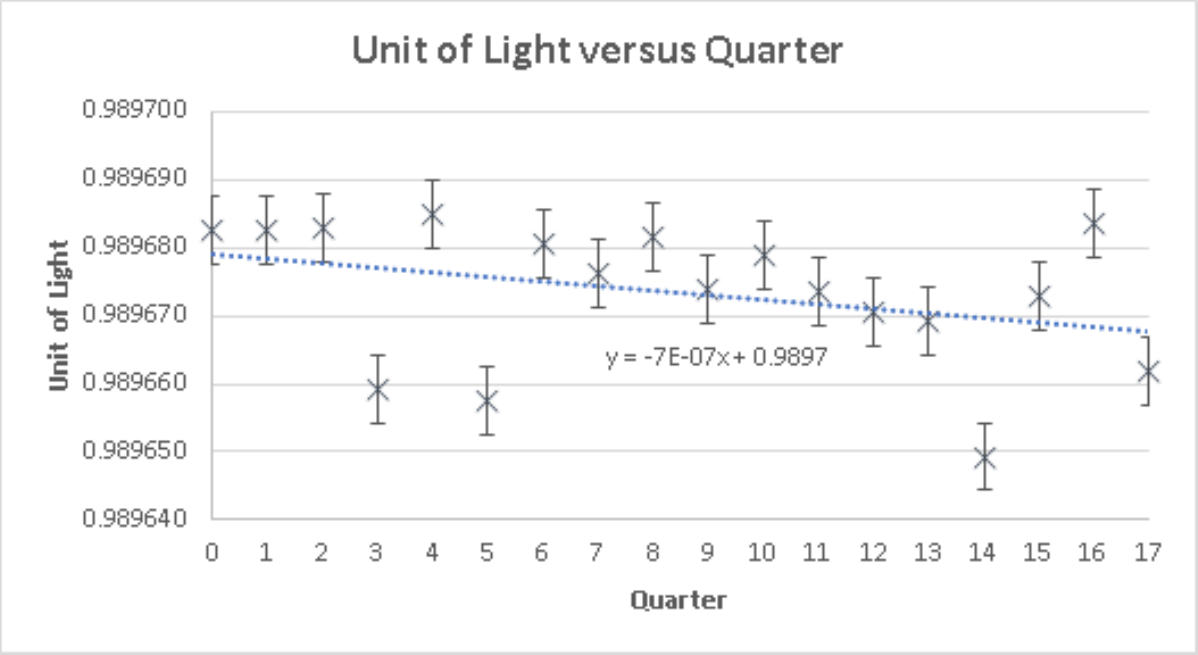} 
\label{fig:K2du}
\caption{Trend of reference light levels (the parameter $U$) over
the 17 quarters of the complete data-set for Kepler 2.}

\end{figure} 
  

\subsection{The Rossiter effect}  

The main points concerning the effect of eclipses of a rotating star on
the measured position of its spectral lines can be summarized (to first
order) by allowing the simplifying assumption that the rotating star is
spherical with radius $R_1$, say, which is significantly larger (by
1/$k$) than that of the planet where $k$ is a small quantity. We
suppose that a spectral line is subject only to a rotational broadening
and not concern ourselves directly with limb-darkening, whose effect can
be separated from that of source motion in the given situation. We then
consider the rotation geometry in relation to a conventional (right) $x,
y, z$ coordinate system with the $z$-axis as the line of sight and the
planet moving across the stellar disk in the direction of positive $x$.

On this basis, the effect of the eclipse of the star, rotating with an
oblique angular velocity vector $\bm{\omega}$, may be interpreted as the
sum of the effects of the three components  of ${\bm \omega}$:
$\omega_x$,  $\omega_y$ and $\omega_z$. The first of these produces a
constant negative or positive net shift (depending on the sign of the
impact factor, with ordinate  $y = b = \cos i/r_1$); the second a shift
that varies linearly from negative to positive or {\em vice versa} for a
retrograde rotation, with central zero; and the third with no RV shift,
but a diminution of the effects of the other two components. In
combination, these effects constitute the basic `Rossiter effect' of
spectral line shifts during the eclipses of rotating stars by exoplanets.

\begin{figure}[t]
\hspace{-0.0cm}
\includegraphics[width=8cm]{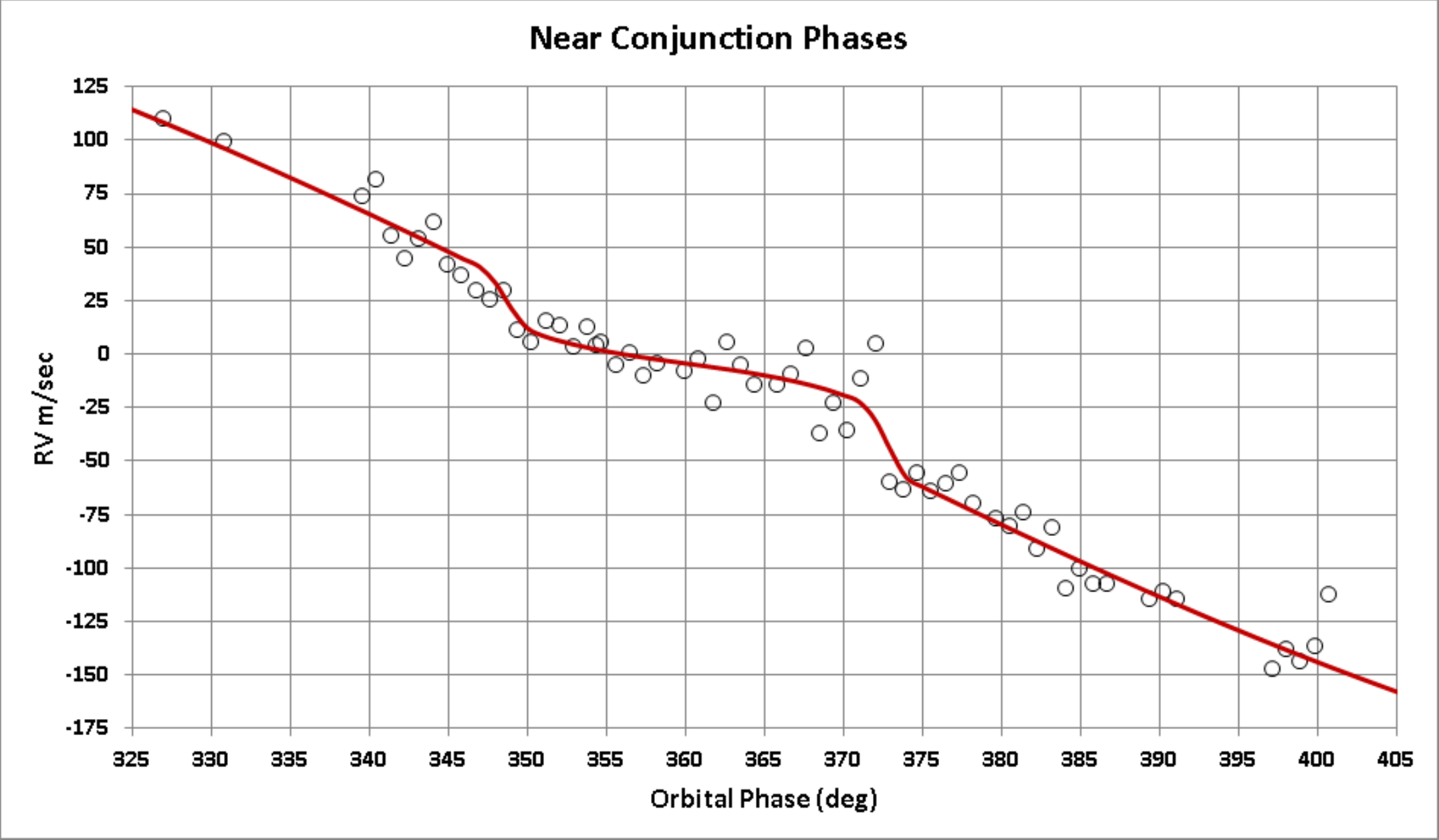} 
\label{fig:Kep2rv}
\caption{Close-up of the near-conjunction phases of the RV curve of Kepler~2A 
using the data-set of (2015).
The eclipse effects on the apparent radial velocities can be seen as:
(1) relatively low in scale (compared with a feasible amplitude of 10s of m s$^{-1}$),
suggesting a fairly high proportion of $\omega_z$ in $\bm{\omega}$; 
(2) inverted, with respect to a positively
rotating star, suggesting a retrograde rotation, i.e.\ $\omega_y$ points in the opposite
direction to ${\bm j}$; (3) although there is significant scatter,
the later upward swing above the orbital trend of the excursion,
 appears greater, in amplitude and extent, than the initial downward swing, 
 suggesting the planet passes over a negative 
 RV associated with $\omega_x$.  The net result is that the 
 conventional `north' rotation pole is in the third quadrant in the observer's 
 $x y$-plane, not far from the disk centre.
}
\end{figure} 

A way to interpret a small amplitude Rossiter effect shown in Fig~6,
then, provided the measured shifts correspond to the movement of the
line centres in the way presented above (cf.\ Hirano et al., 2011;
Bou\'{e} et al., 2013; Brown et al., 2017) is that there is a relatively
large $\bm{z}$ component in $\bm{\omega}$. Unless we have some way to
estimate independently the rotation rate for the star, however, the
projection angle between $\bm{z}$ and $\bm{\omega}$ will have a strong
anti-correlation with that rate.

On the other hand, the scale of the net Rossiter shift at the origin,
compared with its amplitude, relates to the ratio
${\omega_x}/{\omega_y}$.  This appears directly discernible in Fig~6,
and it corresponds to the determinability of the `tilt angle' $\lambda =
\arctan ({\omega_x/\omega_y})$.\footnote{This defines $\lambda$ as a
`position angle', measured clockwise from the $y$-axis. The value of
$\lambda$ measured in this way was found to be about 205$^{\circ} \pm
10^{\circ}$. This is compatible with the angle evaluated by Narita et
al.\ (2009), but their number is $\sim$360$^{\circ}$ less than ours. It
does not agree with the value adopted by Masuda (2015), who gave an
angle less than 180$^{\circ}$, which would place the pole in the fourth
quadrant of the disk at phase zero, unless perhaps an alternative
convention has been followed about the sense of the tilt angle.} In
principle, two angles are involved in locating the position of the
rotation pole from that of the orbit: in practice, the angle $\lambda$
can be estimated directly, whereas the angles $\epsilon$ (obliquity) and
$\psi$ (precession angle)\footnote{In some treatments $\psi$ is used for
the obliquity angle. We have followed an older notation in using
$\epsilon$ (cf.\ e.g.\ Allen, 1974).} tend to show inter-correlation
effects to allow constancy of the amplitude/shift ratio in Fig~6. But,
given the special condition following from the implied closeness of the
line of sight to the disk centre, the angles $\epsilon$ and $\psi$ are
rather tightly constrained to be close to 90$^{\circ}$ ($\epsilon$) or
small ($\psi$). And in that case, the formal derivation of $\lambda$
from $i$, $\epsilon$ and $\psi$ becomes very sensitive to those angle
values. Although this special situation cannot be ruled out from the
analysis thus far, the greater range of parameter space afforded by a
significantly lower stellar rotation rate would tend to push the
modelling towards that more typical geometrical arrangement. 

In fact, Lund et al.\ (2014) argue for a lower value for the rotation
speed, which on the basis of their astero\-seismological evidence turns
out to be close to 7.66 km sec$^{-1}$ at the equator, implying $\gamma_1
=  0.167$.  The average value of $v_{\rm rot} \sin i_{\star}$ from the 6
independent studies listed by Lund et al.\ (2014) is  3.35$\pm$1.11 km
sec$^{-1}$. This would allow $i_{\star}$ to deviate by as much as $\sim
25^{\circ}$ from the line of sight, giving a greater general
probability. Lund et al.\ (2014) cite the comprehensive study of Nielsen
et al.\ (2013) for a comparison of measured stellar rotation periods
which, though showing an appreciable scatter, for their adopted  F6 type
dwarf classification of {\em Kepler}-2 (Feidi et al., 2013) lowers the
value of $\gamma_1$ from that in Table~3 by a factor $\sim$2.  Even so
this would still be twice as fast as Lund et al.'s (2014) rotation rate,
while the value considered by Masuda (2015), in order to produce the
scale of gravity-effect in his modelling of the light curve, required
the fast $\gamma_1 \approx 0.94$.

\begin{table}[t]
\begin{center}
\caption{Results of a follow-up set of
transit model fittings to the complete (binned) PDCSAP data-set of Kepler~2.
note that the displacement of the rotation axis here does not agree with that in 
Table~3.  This is discussed in Section~4.
A quadratic formula for the limb-darkening was adopted. Due to the strong correlation
between $u_1$ and $u_2$ in this formulation the value of $u_1$ is 
significantly larger than its value in Table~2.
 }
\label{tbl-4} 
\begin{tabular}{lrr}
\hline
\multicolumn{1}{c}{Parameter} & \multicolumn{1}{c}{Value} & 
\multicolumn{1}{c}{Error} \\
\hline\\
 $U$           	 & 0.998606       &   0.000003     \\  
 $\Delta \phi_0$ (deg) & 0.007    &   0.008          \\
 $r_1$           & 0.2549         &   0.0001       \\
 $r_2$           & 0.0202         &   0.0001        \\
 $ i $  (deg)    & 81.10          &  0.09           \\
 $u_1$           & 0.553          &  0.005     \\
 $u_2$           &--0.092          &   ---              \\
 $\epsilon$ (deg)  &110.0            &  ---    \\
 $\psi$ (deg)       &16.6          &  0.1     \\   
$\gamma_1$        & 0.245            &  0.06     \\   
$\Delta l$       &0.000009         &  ---               \\
$ \chi^2/\nu$    &1.006            &  ---               \\
\hline
\end{tabular}
\end{center}
\end{table} 


\subsection{Further transit analysis}

The preliminary photometric model characterized by Fig~2 and Table~2 can
be improved by allowing for a non-aligned asynchronous rotation that can
account for light curve asymmetry, together with a more refined
limb-darkening prescription, as with Kepler~13 (Budding et al, 2018). 
This entails adjustment of relevant parameters in the light-curve
modelling, though, as with the RV analysis, geometrical ambiguity may
arise.

With regard to the depression in Fig~2 (lower panel), if this may be
associated with the passage of a planet over a region of surface locally
brightened by a rotationally induced polar flattening, the pole being
not far from the planet's path projected onto the surface, then we could
expect a light variation of order $r_1^3 \gamma_1^2 k^2 \Phi/3$ (Budding
et al.\ 2018), where $\Phi < 1$ is associated with the projection
geometry. In the present case, this implies the maximum scale of a
rotation related anomaly in the transit of {\em Kepler}-2 is about
0.00003, with $\gamma_1 = 1$.  Note that bringing the pole closer to the
line of sight {\em enhances} the effect, unlike the situation for the
Rossiter effect which is {\em reduced} by bringing the polar axis closer
to the line of sight. This suggests combining photometric and
spectroscopic evidence could separate the geometry from the scale of the
rotation speed. Unfortunately, the data does not appear to allow a
self-consistent picture on this, as we see below.

Given the possibility of the lower obliquity proposed by Lund et al.\
(2014), the rotation geometry was set with $\epsilon = 110^{\circ}$ and
the curve-fitting allowed for some adjustment of $\gamma_1$. The
optimization sequence was aimed at removing the small depression at 
phase $\sim0.015$, on the assumption that it was a gravity effect. This
is  essentially similar to the purpose of Masuda (2015), except that the
scale of the anomaly remaining in the transit data used by Masuda (2015)
after removal of his regular eclipse model is about double that of
Fig~2 (0.00001).

Our results, presented in Table~4 and shown in Fig~7, show that it is
possible, with a judicious choice of location for the rotation axis, to
largely remove the depression that appears in Fig~2 when interpreted as
a gravity effect. Moreover, the change of $\gamma_1$  to that of Table~4
can reproduce the the Rossiter effect with essentially the same $\chi^2$
quality of fit as that of Table~3, the obliquity moving to the same
110$^{\circ}$ as in Table~4 by the optimisation process. Note, though,
that the rotation velocity is still $\sim$50\% up on the (uppermost)
value of that of Lund et al.\ (2014), while precession angle $\psi$ is,
in any case, small but positive, so not consistent with the pole
position from the Rossiter effect corresponding to the (2009)
spectroscopic data, which requires $\psi < 0$.

\begin{figure}[t]
\begin{center}
\includegraphics[width=8cm]{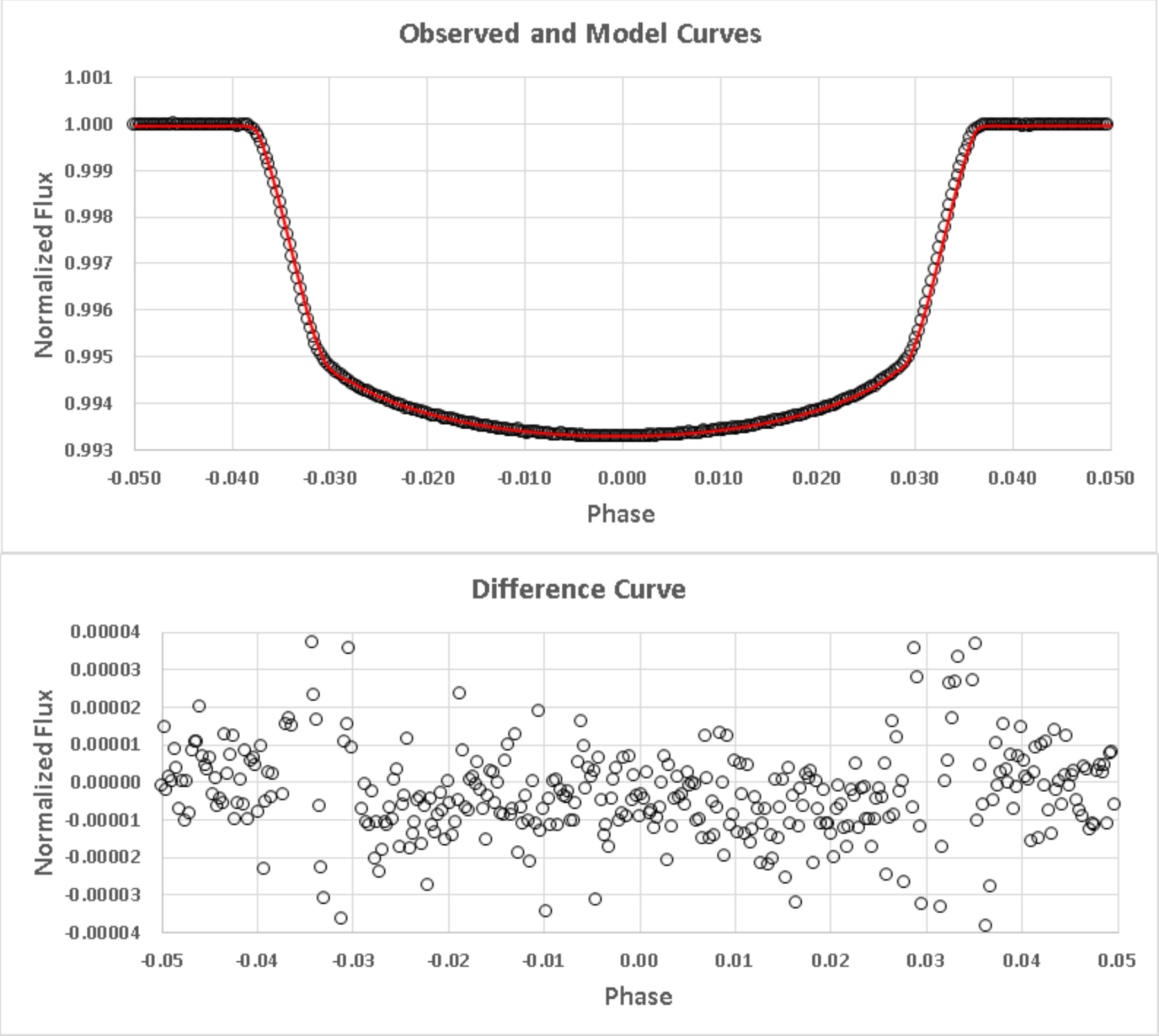} 
\end{center}
\caption{ Fitting to the transit region of the {\em Kepler}-2 light curve, 
using data collected and binned from the full NEA compilation,
with a concentration on the annular region. 
Relocation of the rotation axis towards the centre-right of the visible
disk can remove the depression visible in Fig~2 on the basis that it
is a consequence of gravity-heating of the surface over which the planet passes. 
The partial phases, for which the limb-darkening errors become relatively large, 
are ignored for this context {\color{black} (see Section 3.1)}.  
}
\end{figure}
 
\subsection{Full light-curve fitting and final parameters}
 
\begin{table}[t]
\begin{center}
\caption{Curve fitting results for the complete set of Kepler-2 photometry
 over the whole range of phases.  
\label{tbl-5}} 
\begin{tabular}{lrr}
\hline
\multicolumn{1}{c}{Parameter} & \multicolumn{1}{c}{Value} & 
\multicolumn{1}{c}{Error} \\
\hline\\
$U$ & 1.0000194 & 5$\times 10^{-7}$ \\
$\Delta\phi_0$ (deg) & 0.014 & 0.009\\
$q$ & 0.00101 & 0.00042\\
$\tau_1$ & 0.87 & 0.01\\
$E_2$ & 0.202 & 0.003 \\
$\Delta l $& 0.000009 & --- \\
$\chi^2/\nu$ & 1.12 & ---\\
\hline
\end{tabular}
\end{center}
\end{table}
   
The full perturbation of the light from an idealised spheroidal model
can be regarded as made up of 3 terms (Budding et al., 2018; see also
Shporer et al., 2011; Faigler et al., 2013; and other historic papers on
close binary light curves) arising from the tidal distortion  (dependent
on the mass-ratio $q$ and gravity-effect $\tau_1$), reflection effect
from the companion object (coefficient $E_2$) and the Doppler effect
from the star's orbital motion ($K_1$); thus, essentially:
 \begin{equation} 
 \Delta L = \Delta L_{q,{\tau_1}} + \Delta L_{E_2} + \Delta L_{K_1} \,\,\, . 
 \end{equation} 
 
Formula (4) is strictly valid only through the uneclipsed phases. During
eclipses the $\Delta L$s are replaced by expressions of the form
$(\Delta - \delta)L$, where $\delta$ indicates that some proportion of
the light perturbation $\Delta$ is eclipsed out. In practice, the
$\Delta L_{q,{\tau_1}}$ term is very small in the planetary transit
phases; and so is $\Delta L_{E_2}$, although, clearly with a visible
occultation, the planet's light loss is that of its reflected light. The
Doppler term becomes noticeable in the transit if the star's rotation is
sufficiently  high, leading to a photometric parallel with the
spectroscopic Rossiter effect discussed in the previous subsection. This
can be shown to be of order $\alpha V_{\rm rot} \sin i_{\star} k^2 /c$
(Shporer et al., 2012), where $\alpha$ is a wavelength and temperature
related coefficient of order unity, similar to that affecting $\tau$ in
Section~3.1. The velocity ratio is already of order 10$^{-5}$; when we
multiply by the other small factors the contribution becomes
insignificant in the present case. 
 
\begin{figure}[t]
\includegraphics[width=8cm]{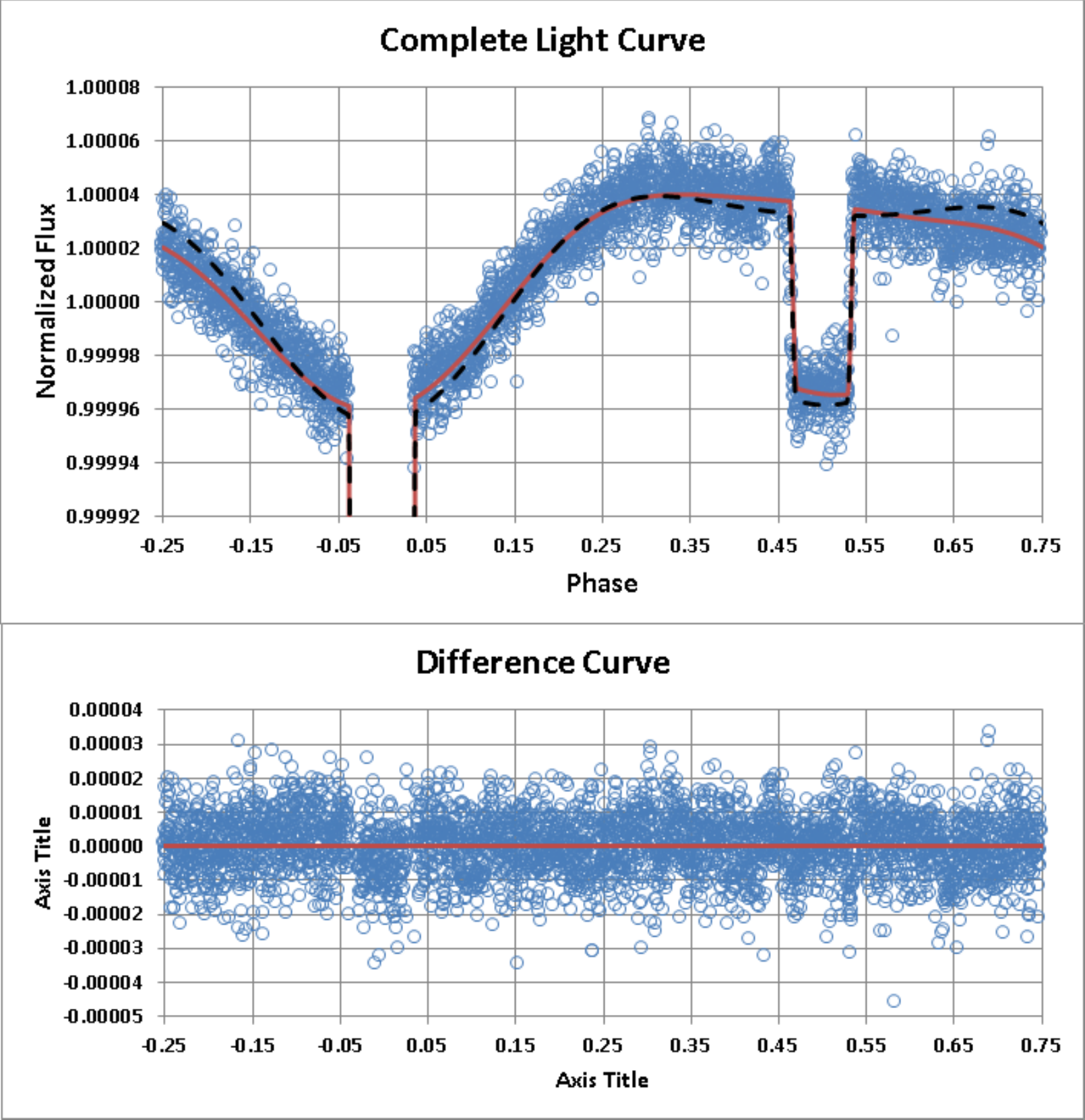} 
\vspace{-0.5cm}
\label{fig:Kepflc}
\caption{ The complete light curve is shown with corresponding residuals (lower panel). The 
optimal model's direct tidal, reflection and Doppler effects can be seen (dashed).
The better-fitting curve (continuous) comes from a slight `cleaning' of the model (Section~4 of text). 
}
\end{figure} 

The coefficients of these three terms in the out-of-eclipse regions
might be regarded as adjustable parameters (e.g. Lillo-Box et al.,
2014). In practice, the scale of the minute $\sim10^{-6}$ orbital
Doppler effect should be specified by the prior mass $M_{\star}$ and the
already confirmed mass ratio $q$ and orbital inclination $i$ that cannot
be too far from 90$^{\circ}$ for eclipses to occur. Allowing  $\Delta
L_{K_1}$ to become a separate free parameter raises physical
inconsistency as well as possible mathematical determinacy issues
(Barclay et al., 2015). In our findings, however, the apparent Doppler
effect in the light curve appears significantly greater than its
theoretically predicted value.  This point was already noticed by
Esteves et al.\ (2013) and Faigler  \& Mazeh (2015).

Our results of fitting the full light curve are given in Table~5, where
optimal fittings to the binned data concerned the mean reference flux
$U$, the mass-ratio $q$ and its closely correlated gravity-parameter
$\tau_1$, together with the reflection coefficient  $E_2$; the other
parameters having been fixed from previous fittings. We can regard $E_2$
as proportional to the geometric albedo $A_g$ for the limiting case of a
full radiator obeying Lambert's law (Budding et al., 2018). Allowing
$E_2$ to depart from this ideal case as a scaling factor for the
returned light involves the same kind of interpretation as having an
empirical $A_g$ that differs from the Bond albedo in dependence on the
surface properties.  Thus a high $E_2$ value is suggestive of clouds or
efficient light-scattering atmospheric particles.  In fact, optimal
values from curve-fitting experiments remained close to the luminous
efficiency formula of Hosokawa (1958), for which  $E_2 \approx 0.25$.
This is an order of magnitude higher than the corresponding coefficient
for {\em Kepler}-1 (Budding et al., 2018).

The fitting for the complete light curve has a comparable overall
quality of fit  ($\chi^2/\nu$) to that of the transit region of Table~4
with the adopted datum error of 9 ppm, but that is still somewhat higher
than the 7 ppm that could be expected from the raw measurements if the
residuals were due only to Poissonian photon noise (Section 3.1). The
derived mass ratio is slightly less than that from Winn et al.\ (2009).
The finally adopted light curve model and residuals are shown in Fig~8.

There is a small difference between the best-fitting unaligned
asynchronous {\sc WinFitter} model (dashed curve) and a fitting 
corresponding to an empirically `cleaned' model (continuous curve). In
this the residuals from the original light curve were fitted with a
low-order Fourier decomposition consisting of only the constant, phase
and 2$\times$phase terms, i.e.\
\begin{equation}
 \Delta L_{\rm res} = a_0 + a_1 \cos \phi + b_1 \sin \phi + a_2 \cos 2 \phi + b_2 \sin 2 \phi  .
\end{equation}
The procedure resembles that of Zeilik et al.\ (1988) in the context of
distorted close binary system light curves. The results are presented in
Table~6.

\begin{table}[t]
\begin{center}
\caption{Curve fitting the complete set of Kepler-2 photometry
 over the whole range of phases: empirical adjustment of the leading terms.  
\label{tbl-5}} 
\begin{tabular}{lrr}
\hline
\multicolumn{1}{c}{Coefficient} & \multicolumn{1}{c}{Value (ppm)} & 
\multicolumn{1}{c}{Error} \\
\hline\\
$a_0$ & 0.1 & 0.5 \\
$a_1$ & 0.0 & 0.7\\
$b_1$ & 4.2 & 0.7\\
$a_2$ & 4.4 & 0.7 \\
$b_2$ & --0.8 & 0.7 \\
$\Delta l $& 0.9 & --- \\
$\chi^2/\nu$ & 0.93 & ---\\
\hline
\end{tabular}
\end{center}
\end{table}

Given the very low scale and separability of the Fourier terms we may
regard $b_1$ as the above-mentioned increase on the $\Delta L_{K_1}$
contribution and $a_2$ as a slight decrease on the tidal term.  These
points are considered below.

  
\section{Summary and discussion}
  
Table~7 lists our adopted set of absolute parameters, calculated from
taking into account the results of the analyses in Tables~1 to 5,
together with corresponding error estimates. The numbers in Table~7,
that  depend on separately determined and often less accurately known
priors, are shown with the corresponding parameters from the NEA
archive, supplemented with other literature data mentioned in Section
2.2. The two sets of results are reasonably compatible when compared
with the error estimates, though our findings are for somewhat larger,
less condensed, bodies.

\begin{table}[t]
\begin{center}
\caption{Reference parameters: derived and literature comparisons.
{\color{black} Conversion factors were calculated using values from Allen (1973), such as for the solar radius.}
\label{tbl-5}} 
\begin{tabular}{lccc}
\hline
\multicolumn{1}{c}{Parameter} & \multicolumn{1}{c}{Value} &
\multicolumn{1}{c}{Error} & \multicolumn{1}{c}{Lit.} \\
\hline\\
$M_{\star}$ \small{ $\odot$ } & 1.53 & 0.09 & 1.53 \\
$M_p$ \small{Jup} & 1.80 & 0.13 & 1.67\\
$a$ km$\times 10^6$ & 5.72 & 0.19 & 5.50 \\
$R_{\star}$ km$\times 10^6$ & 1.46 & 0.08 & 1.36 \\
$R_p$ km$\times 10^5$ & 1.15 & 0.07 & 1.03 \\
$\rho_{\star}$ SI$\times 10^{-3}$ & 0.24 & 0.04 & 0.29 \\
$\rho_p$  SI$\times 10^{-3}$      & 0.48 & 0.06  & 0.56 \\
$A_g$ & 0.25 & 0.02 & 0.27 \\
\hline
\end{tabular}
\end{center}
\end{table} 

Previous discussions of the ellipticity $\Delta L_{q,{\tau_1}}$ tend to
be limited to the principle ($-\cos 2\phi$) term and taken to be
independent of the reflection ($\sim \cos \phi$) and Doppler ($\sin
\phi$) effects. We could therefore expect our more detailed formulation
to give a more accurate result.  But this possibility of high accuracy
materializes only if there are no other factors in the data than those
allowed for in the model. Fig~8 (upper panel) shows that this is not
quite the case.

In fact, by adding the low-order Fourier components of Table~6 into the
fitting function, the $\sin \phi$ term increases its amplitude to
$\sim$5 ppm. But this is $\sim$6 times the scale of the Doppler-effect
amplitude that can be accurately specified from the $K_1$ value of
Table~3. The conclusion must then be that there are other contributing
factors to the light-curve not included in the model.  This seems
self-evident for the asymmetric term in $\sin \phi$.

Fig~8 shows that the increase in the $\sin \phi$ term in the first half
of the light curve closely matches the decrease in the leading
ellipticity term there; whereas in the second half of the curve the two
decreases support each other, making the net asymmetry of the light
curve significantly greater than in the standard model. The tidal and
reflection terms are to some extent correlated, since there is also a
tidal term in $\cos \phi$.  However, from Table~6 we deduce that any
such residual effects are not so significant against the scale of
measurement error.

The sense of the persisting asymmetry in Fig~8 is different from that of
{\em Kepler}-1, where it is very small but in reasonable agreement with
the model, and {\em Kepler}-13, where the rise in the second half of the
light cycle over-compensates for the Doppler effect and brings the two
observed maxima to almost the same level.  This latter situation agrees
with the clouding scenario of von Paris et al.\ (2016), in which the
`morning' (western, approaching) limb is brightened when  westerly winds
spill over beyond the night-time terminator, producing increased cloud
and brightening the approaching planet.  For {\em Kepler}-2 the
prevailing tendency seems to be in the opposite direction, according to
Fig~8, conforming more with the  eastward hot-spot shift model of
Faigler \& Mazeh (2015), in which the afternoon (eastern, receding) limb
becomes relatively bright. The difference between these otherwise
generally comparable hot jupiters on this matter lends support to the
fluctuating surface-weather discussion of Armstrong et al.\ (2016).

In Fig~8 we see that the light level at the centre of the planetary
occultation is closely similar to that just outside the transit eclipse,
tending to reinforce the scenario referred to by Borucki et al. (2009)
and Wong et al.\ (2016) of limited heat transfer between the illuminated
and non-illuminated hemispheres of the planet. But the irregularity in
the light curve residuals around the primary minimum region, of order a
few ppm, suggest that the Lambertian reflection model may not be
adequate to cover the physics of the situation. The model adopts that
the inherent light of the planet is zero, but the anomaly about the
transit suggests the possibility of a low back\-warming effect from the
heated surface of the planet.

The planet has a Safronov parameter ($\Theta = q/r_2 = 0.049 \pm 0.008$;
Safronov, 1972) and therefore within the `Class I' grouping of the more
massive hot jupiters. Hansen \& Barman (2007) suggested that $\Theta$
should be connected to the equilibrium temperature of a planet in view
of atmospheric evaporation tending to decrease $q$ and increase $r_2$.
This inference may be connected to the moving bright spots or strong
winds discussed by Borucki et al.\ (2009), Esteves et al.\ (2015),
Armstrong et al.\ (2016) and others.  Stability of the outer regions of
hot jupiters continues to stimulate lively discussion (Christian \&
Lund, 2010; Madhusudhan et al., 2018).

In Section 3 we noted that there was a disparity in the location of the
polar axis between the spectroscopic and photometric analyses. There are
several lines of thought that bear on this. Firstly, although the pole
is located in different quadrants of the disk at phase zero in these
different findings, there is a general agreement  (also with Winn et
al., 2009), that the rotation axis is not far from the line of sight
(see also, Southworth, 2017). Our spectroscopic analysis referred
to the relatively high degree of scatter in the transit-phase data,
though the separation of the two pole positions is greater than
reasonable error estimates would allow their coincidence. Adopting,
then, that these two analyses do result in essentially different axis
positions, there arise such explanation possibilities as (a) the
spectroscopic or photometric modelling retains some inadequacy, or (b)
there has been a measurable movement of the pole between mid-2008 (the
Subaru observations) and mid-2011 (middle of the Kepler series).

Concerning point(a) and in relation  to the spectroscopic results, it
can be kept in mind that the conventional model adopted is that the
effect of a transiting planet on a spectral line's Doppler shift arises
from a convolution  with other processes, particularly instrumental
broadening. The measured movement of the line is then that of its `light
centre' (Hirano et al., 2011; Brown et al., 2017). If, with a very high
resolution system, a line profile broadened only by rotation were to be
recorded there would be no Rossiter effect on the centre of the line.
The eclipsing planet would simply register as a feature on the
profile.\footnote{This is the basis of the `Doppler imaging' technique
(cf.\ Vogt, 1983).} The HDS data shown in Fig 7 is from a high
resolution spectrograph, but it is not clear to what extent this may
bear on the data-processing used to determine reported wavelength
shifts.

Similarly, with regard to the photometry, stars may be affected by other
local darkening or brightening mechanisms than the gravity effect. Even
within the physics of that effect there is scope for departure from the
{\sc WinFitter} specification of rigid body rotation, for example, which
stars do not strictly conform to.  At the same time, stellar circulation
currents may counter the classic von Zeipel formulation of this effect
(K{\i}rb{\i}y{\i}k \& Smith, 1976). Also, the mid-F type classification
of the host star suggests the possibility of a `polar spot' (Schrijver~\&
Title, 2001; Berdyugina, 2002; Oshagh et al., 2015). Instead of the
depression in Fig~2 being due to the eclipse of a bright region, it
could be that the raised part of the residuals distribution in the
minimum is due to the eclipse of a somewhat cooler part of the surface.
Other modelling or data inadequacies were mentioned in Section 2.2.
Having said this, the fairly low extent of model inadequacy relative to
data scatter can be judged by the  difference between the continuous and
dashed lines in Fig~8.

Point (b) requires a movement of several deg in $\sim$3 y, i.e.\ a
precession period of order a few hundred y only. This is about an order
of magnitude less than could be reasonably expected for the
configuration (see, for example, the calculation for the comparable case
of $Kepler$-13 in Budding et al., 2018). It seems, then, much more
likely that the apparent shifts of axis position between the 2008 and
2011 (averaged) data-sets are related to inadequacy of modelling for the
spectroscopic Rossiter effect, or additional, unprogrammed, surface
inhomogeneities of brightness affecting the photometric analysis, than
real short-term movements of the star's rotation axis.  Continued
detailed monitoring of {\em Kepler}-2 (HAT-P-7A) are required to help
solve these remaining problems of the system.


\section {Acknowledgements}

It is a pleasure to thank Prof.\ Osman Demircan and the colleagues in
the Physics Department of COMU (\c{C}anakkale, Turkey) for their
interest and support of this programme. The research has been supported
by T\"{U}B\.{I}TAK (Scientific and Technological Research Council of
Turkey) under Grant No.\ 113F353. Additional help and encouragement for
this work has come from the National University of Singapore,
particularly through Prof.\ Lim Tiong Wee of the Department of
Statistics and Applied Probability.  This research has made use of the
NASA Exoplanet Archive, which is operated by the California Institute of
Technology, under contract with the National Aeronautics and Space
Administration under the Exoplanet Exploration Program. {\color{black} 
This work has made use of data from the European Space Agency (ESA) mission
{\it Gaia} (\url{https://www.cosmos.esa.int/gaia}), processed by the {\it Gaia} Data Processing and Analysis Consortium (DPAC,
\url{https://www.cosmos.esa.int/web/gaia/dpac/consortium}). TB \&
EB wish to record our gratitude to the late Prof.\ Denis J.\ Sullivan (Victoria University of
Wellington, NZ) for his support of this and related programmes, and extend our condolences
to his family.}

\end{document}